\documentclass[prl,aps,twocolumn,preprintnumbers,amsmath,amssymb,superscriptaddress,showpacs]{revtex4-1}

\usepackage{amsmath}
\usepackage{graphicx}
\usepackage{amssymb}
\usepackage{color}
\usepackage[dvipdfm,bookmarks=true,colorlinks=true,linkcolor=blue,urlcolor=blue,citecolor=blue]{hyperref}

\newcommand{\be}{\begin{equation}}
\newcommand{\ee}{\end{equation}}

\newcommand{\bit}{\begin{itemize}}
\newcommand{\eit}{\end{itemize}}
\newcommand{\bea}{\begin{eqnarray}}
\newcommand{\eea}{\end{eqnarray}}

\usepackage{epsfig}

\begin{document}

\title
{Gapless spin-liquid ground state in the $S = 1/2$ kagome antiferromagnet}

\author{H. J. Liao}
\affiliation
{Institute of Physics, Chinese Academy of Sciences, P.O. Box 603, Beijing
100190, China}

\author{Z. Y. Xie}
\affiliation
{Institute of Physics, Chinese Academy of Sciences, P.O. Box 603, Beijing
100190, China}
\affiliation
{Department of Physics, Renmin University of China, Beijing 100872, China}

\author{J. Chen}
\affiliation
{Institute of Physics, Chinese Academy of Sciences, P.O. Box 603, Beijing
100190, China}

\author{Z. Y. Liu}
\affiliation
{Institute of Theoretical Physics, Chinese Academy of Sciences, P.O. Box
2735, Beijing 100190, China}

\author{H. D. Xie}
\affiliation
{Institute of Physics, Chinese Academy of Sciences, P.O. Box 603, Beijing
100190, China}

\author{R. Z. Huang}
\affiliation
{Institute of Physics, Chinese Academy of Sciences, P.O. Box 603, Beijing
100190, China}

\author{B. Normand}
\affiliation{Laboratory for Neutron Scattering and Imaging, Paul Scherrer
Institute, CH-5232 Villigen PSI, Switzerland}
\affiliation
{Department of Physics, Renmin University of China, Beijing 100872, China}

\author{T. Xiang}\email{txiang@iphy.ac.cn}
\affiliation
{Institute of Physics, Chinese Academy of Sciences, P.O. Box 603, Beijing
100190, China}
\affiliation{Collaborative Innovation Center of Quantum Matter, Beijing
100190, China}

\date{\today}

\begin{abstract}
The defining problem in frustrated quantum magnetism, the ground state
of the nearest-neighbor $S = 1/2$ antiferromagnetic Heisenberg model on
the kagome lattice, has defied all theoretical and numerical methods
employed to date. We apply the formalism of tensor-network states (TNS),
specifically the method of projected entangled simplex states (PESS),
which combines infinite system size with a correct accounting for
multipartite entanglement. By studying the ground-state energy, the
finite magnetic order appearing at finite tensor bond dimensions, and the
effects of a second-neighbor coupling, we demonstrate that the ground state
is a gapless spin liquid. We discuss the comparison with other numerical
studies and the physical interpretation of this result.
\end{abstract}

\maketitle

In one spatial dimension (1D), quantum fluctuations dominate any physical
system and semiclassical order is destroyed. In higher dimensions, frustrated
quantum magnets offer perhaps the cleanest systems for seeking the same
physics, including quantum spin-liquid states, fractionalized spin degrees
of freedom, and exotic topological properties. This challenge has now become
a central focus of efforts spanning theory, numerics, experiment, and
materials synthesis \cite{lbn,qslkagome,qslto,rn}. While much has been
understood about frustrated systems on the triangular, pyrochlore,
Shastry-Sutherland, and other 2D and 3D lattices, it is fair to say that
the ground-state properties of the $S = 1/2$ kagome Heisenberg antiferromagnet
(KHAF) remain a complete enigma.

An analytical Schwinger-boson approach \cite{rs}, coupled-cluster
methods \cite{rgfblr}, and density-matrix renormalization-group (DMRG)
calculations \cite{rjws,ryhw,rdms}, including analysis of the topological
entanglement entropy \cite{rjwb}, all suggest a gapped spin liquid of Z$_2$
topology. The most sophisticated DMRG studies \cite{rdms,rnsh} estimate a
triplet spin gap $\Delta \ge 0.05 J$. Analytical large-$N$ expansions
\cite{rrhlw} and numerical simulations by variational Monte Carlo (VMC)
\cite{ribsp,ripb} suggest a gapless spin liquid with U(1) symmetry and a
Dirac spectrum of spinons. Extensive exact-diagonalization calculations
conclude that the accessible system sizes are simply too small to judge
\cite{rsl,rlsm}. Debate continues between the gapped Z$_2$ and gapless U(1)
scenarios, with very recent arguments in support of both \cite{rlt,ripbc},
while a symmetry-preserving TNS study favors the gapped Z$_2$ ground state
\cite{rmchw}. Experimental approaches to the kagome conundrum have made
considerable progress in recent years, but for the purposes of the current
theoretical analysis we defer a review to Sec.~SI of the Supplementary
Material (SM) \cite{sm}.

In this Letter, we employ the PESS description of the entangled many-body
ground state to compute the properties of the KHAF. Because we consider an
infinite system, our results provide hitherto unavailable insight. As
functions of the finite tensor bond dimension, we find algebraic convergence
of the ground-state energy and algebraic vanishing of a finite staggered
magnetization, indicating a gapless spin liquid. We demonstrate that the
phase diagram in the presence of next-neighbor coupling contains a finite
region of this spin-liquid phase. Our results suggest that the physics of
the KHAF is driven by maximizing the kinetic energy of gapless Dirac spinons.

The TNS formalism is based on expressing the wave function as a generalized
matrix-product state (MPS) \cite{Nig,Nishino,Cirac}. As we review in Sec.~SII
of the SM \cite{sm}, this Ansatz obeys the area law of entanglement and,
crucially, allows the construction of a renormalization-group scheme to reach
the limit of infinite lattice size. The truncation parameter is the tensor
bond dimension, $D$. We introduced the PESS formulation \cite{XCY+14} in order
to capture the multipartite entanglement within each lattice unit, or simplex
\cite{Arovas08,XCY+14,LXC+16}, which is the key element of frustrated systems
and is missing in the conventional pairwise projected entangled pair states
(PEPS) construction. Summarizing the numerical procedure (Sec.~SII \cite{sm}),
the optimized PESS approach is a projection technique, with tensor manipulation
performed by higher-order singular-value decomposition (HOSVD), and freedom to
choose the simplex, the unit cell, and a simple- or full-update treatment of
the bond environment during tensor renormalization, the former allowing access
to larger $D$ but the latter achieving more rapid convergence.

\begin{figure}[t]
\begin{center}
\epsfig{file=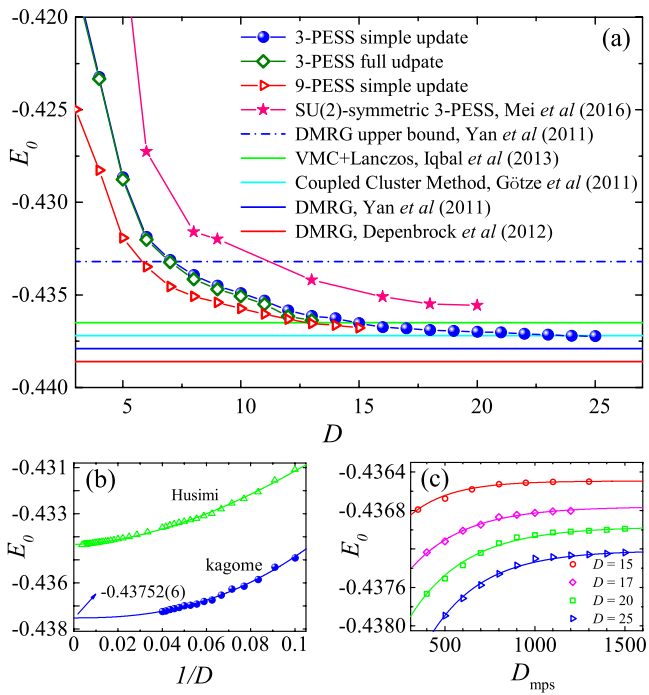,width=8.5cm,angle=0.0}
\end{center}
\caption{{\bf Ground-state energy of the KHAF.} (a) $E_0$ as a function of
$D$, shown for the 3-PESS and simple-update method up to $D = 25$, 3-PESS
by full update to $D = 13$, and 9-PESS with simple update to $D = 15$. Shown
for comparison are results from other numerical studies. (b) $E_0(D)$ for
the 3-PESS, shown as a function of $1/D$ and compared with results obtained
for the Husimi lattice \protect{\cite{LXC+16}}. (c) Convergence of $E_0(D)$
as a function of $D_{\rm mps}$, shown for several values of $D$.}
\label{F1}
\end{figure}

However, TNS calculations are a two-step process, where the wave function
is obtained first and then used to calculate physical expectation values.
This latter step requires projection onto a 1D MPS basis, whose dimension for
convergence is found to scale approximately as $D_{\rm mps} \approx 4 D^2$.
Once $D \gtrsim 15$, the evaluation step becomes the more computationally
intensive problem, and here we implement new methodology (outlined in
Sec.~SII \cite{sm}) by which we extend the accessible $D$ range.

We begin by presenting results from the 3-site-simplex (3-PESS) Ansatz
for all accessible $D$ values. The ground-state energy, $E_0(D)$, of the
nearest-neighbor KHAF is shown in Fig.~\ref{F1}(a). At large $D$, our estimate
lies below those obtained from all known techniques other than DMRG studies of
specific clusters, which are not an upper bound. We remark that our $E_0(D)$
values are significantly lower than those of an SU(2)-invariant TNS analysis
\cite{rmchw}. We find that $E_0(D)$ converges algebraically with $D$, as
on the Husimi lattice \cite{LXC+16}, indicating a gapless ground state
\cite{rpvvt}. The power-law form $E_0 (D) = e_0 + a D^{-\alpha}$, shown in
Fig.~\ref{F1}(b), delivers our best estimate of the ground-state energy,
$e_0 = - 0.43752(6) J$. Figure \ref{F1}(c) illustrates the convergence of
$E_0(D_{\rm mps})$ for selected values of $D$; we note that this part of the
process is not variational and comment in detail in Sec.~SII of the SM
\cite{sm}. Optimized fits to a regime of exponential convergence in
$D_{\rm mps}$ were used to extrapolate towards the values of $E_0(D)$ shown
in Figs.~\ref{F1}(a) and \ref{F1}(b), and to determine the associated error
bars, on the basis of which we limit our claims of reliability to $D \le 25$.

\begin{figure}[t]
\begin{center}
\epsfig{file=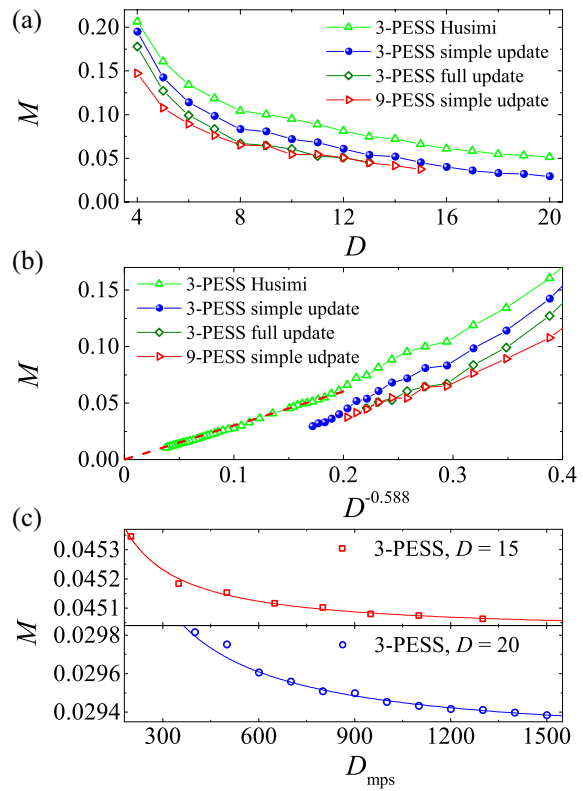,width=7.5cm,angle=0.0}
\end{center}
\caption{{\bf Staggered magnetization of the KHAF at finite $D$.} (a) $M$
as a function of $D$, shown for the 3-PESS and simple-update method up to
$D = 20$, 3-PESS by full update to $D = 13$, and 9-PESS with simple update
to $D = 15$. Shown for comparison are results obtained for the Husimi
lattice \protect{\cite{LXC+16}}. (b) $M$ as a function of $1/D^{0.588}$, the
power-law form obtained for the Husimi lattice. (c) Convergence of $M(D)$
as a function of $D_{\rm mps}$, shown for $D = 15$ and $D = 20$.}
\label{F2}
\end{figure}

One key qualitative property of our PESS wave function is a finite
120$^{\rm o}$ magnetic order at all finite $D$ values, as shown in
Figs.~\ref{F2}(a) and \ref{F2}(b). The order parameter, $M(D)$, varies
algebraically with $1/D$ over the available $D$ range, tending to zero as
$D \rightarrow \infty$, as required of a spin liquid. Figure \ref{F2}(c)
illustrates the convergence of $M(D_{\rm mps})$ for $D = 15$ and 20, where
an algebraic form was deduced from the truncation error, and reliable
extrapolations to large $D_{\rm mps}$ were obtained only for $D \le 20$.

\begin{figure*}[t]
\begin{center}
\epsfig{file=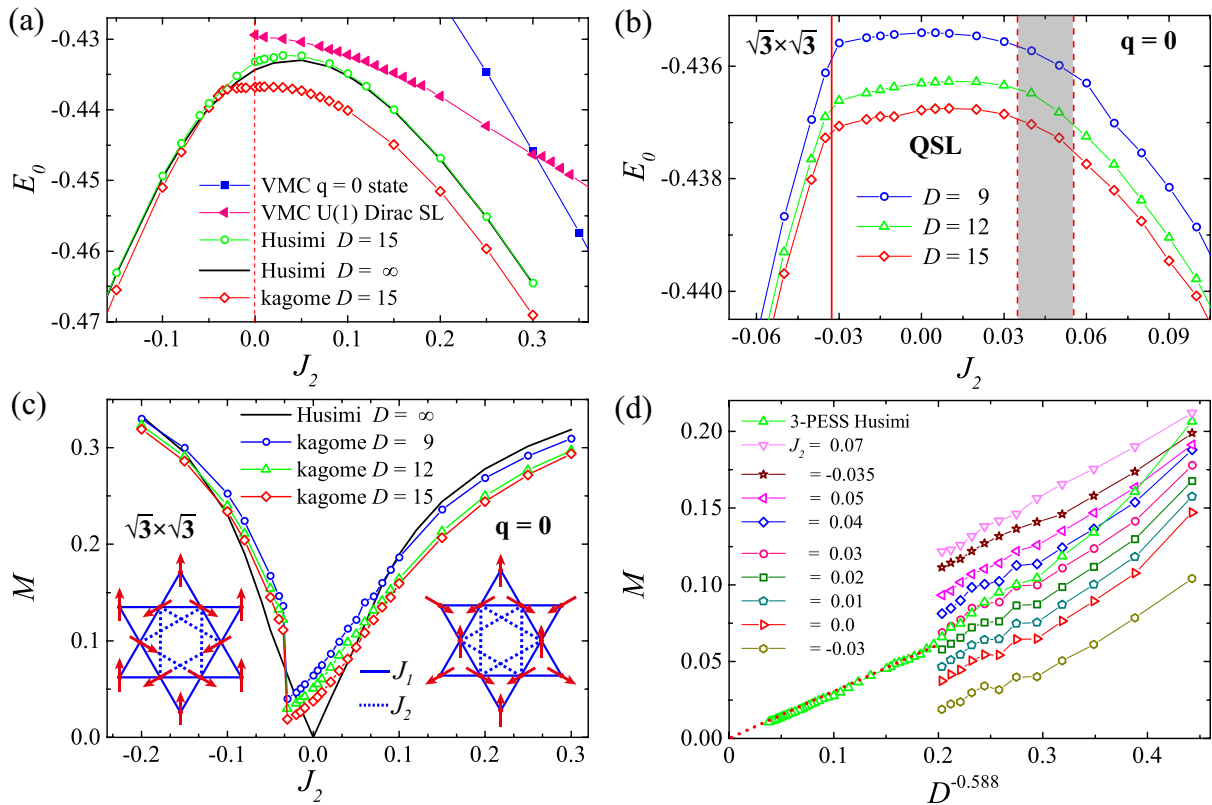,width=16cm,angle=0.0}
\end{center}
\caption{{\bf Energy and staggered magnetization of the KHAF with
next-neighbor coupling.} (a) $E_0(J_2)$ calculated with a 9-PESS using
$D = 15$. Shown for comparison are results for the Husimi lattice with
$D = 15$ and $D = \infty$, as well as the VMC results of
Ref.~\protect{\cite{ripb}}. (b) Detail of $E_0(J_2)$ near $J_2 = 0$; with
increasing $D$, a cusp-type discontinuity emerges near $J_2 = - 0.03$ (red
solid line). The shaded region denotes the location of the continuous
transition at small positive $J_2$, deduced from the magnetization of
panel (d). (c) $M(J_2)$ calculated using $D = 9$, 12, and 15, compared
with results for the Husimi lattice at $D = \infty$. Insets represent the
$\sqrt{3} \times \! \sqrt{3}$ (left) and $q = 0$ ordered phases (right).
(d) $M$ as a function of $1/D^{0.588}$, with the Husimi result indicating
$J_2$ values for which $M$ may extrapolate to zero within the error bars.}
\label{F3}
\end{figure*}

The Husimi lattice provides essential confirmation of our results. It
possesses the same local physics as the kagome lattice, but less frustration
from longer paths, and it allows PESS calculations up to $D = 260$, yielding
accurate extrapolations to the large-$D$ limit \cite{LXC+16}. It confirms the
crucial qualitative statement that magnetically ordered states have the lowest
energies for spatially infinite systems at finite $D$. It benchmarks the
algebraic nature of $E_0(D)$ [Fig.~\ref{F1}(b)] and $M(D)$ [Fig.~\ref{F2}(b)],
the latter vanishing exactly at large $D$. The Husimi $M(D)$ sets an upper
bound on the kagome $M(D)$ (Sec.~SIII of the SM \cite{sm}). Our kagome results
lie well below this bound, but with no evidence for deviation from a similar
algebraic form, reinforcing the conclusion that the ground state of the KHAF
is a gapless spin liquid.

For full rigor we consider every aspect of the PESS procedure. Full-update
calculations confirm the accuracy of the simple-update approximation for all
accessible $D$ values. $E_0(D)$ lies only slightly lower [Fig.~\ref{F1}(a)],
with no change in functional form; similarly, $M(D)$ is suppressed by several
percent [Figs.~\ref{F2}(a) and \ref{F2}(b)], reinforcing the argument for
convergence to $M = 0$ at large $D$, but still shows algebraic behavior. To
investigate whether magnetic order might be artificially enhanced by the
3-PESS, in Figs.~\ref{F1}(a), \ref{F2}(a), and \ref{F2}(b) we also present
results obtained using a 9-site simplex (9-PESS) \cite{XCY+14}, which again
confirm the algebraic form of $E_0(D)$ and $M(D)$, with no evidence either of
a crossover to exponential behavior of $E_0(D)$ or of a collapse of $M(D)$ to
zero at finite $D$. 3-PESS calculations may be performed with a unit cell
containing any number of simplices (Sec.~SII \cite{sm}); our results for 3-,
9-, and 12-site unit cells are identical, again confirming no inherent
bias of this type.

Further essential confirmation is obtained by adding a next-neighbor
coupling, $J_2$. These calculations are performed most efficiently with a
9-PESS and we reach $D = 15$ with simple updates. As shown in Fig.~\ref{F3}(a),
$E_0(J_2)$ is maximal (the system is most frustrated) close to $J_2 = 0$
and is not symmetrical about this point. For the Husimi lattice, $E_0(J_2)$
is continuous, with maximal frustration at $J_2 \simeq 0.04$. By contrast,
the kagome case shows a regime of almost constant energy when $-0.03
\lesssim J_2 \lesssim 0.04$ [Fig.~\ref{F3}(b)]. To understand the nature
of these states, we consider in Fig.~\ref{F3}(c) the finite-$D$ magnetization
and in Fig.~\ref{F3}(d) $M(D)$ for selected values of $J_2$. For the Husimi
lattice, $M(J_2)$ is zero only at $J_2 = 0$, where it has a discontinuity,
and [despite the form of $E_0(J_2)$] is almost symmetrical. For kagome, the
expected ordered phases are the $q = 0$ structure at $J_2 > 0$ and the
$\sqrt{3} \times \! \sqrt{3}$ structure at $J_2 < 0$ [Fig.~\ref{F3}(c)].
However, $M(J_2)$ at finite $D$ continues to fall through $J_2 = 0$ from
above, indicating a region of $q = 0$ order at $J_2 < 0$, which is terminated
at $J_2 \simeq - 0.03$ by a discontinuous jump to $\sqrt{3} \times \!
\sqrt{3}$ order. From Fig.~\ref{F3}(d), $M(D)$ appears to extrapolate to
zero over a range of $J_2$ values, which we estimate from the Husimi
magnetization to be fully consistent with the ``plateau'' in $E(J_2)$
[Fig.~\ref{F3}(b)].

The gapless spin liquid should exist over a finite range of $J_2$ if it is
a robust quantum ground state. The Husimi case, with magnetic order at all
finite values of $|J_2|$ and a spin liquid only at the single point $J_2 = 0$,
is a type of ``phase diagram'' allowed only because of the pathological Husimi
geometry. In the kagome case, indeed we find a finite disordered regime,
bounded by a first-order transition at $J_2 \simeq - 0.03$ and an apparent
second-order transition at $J_2 = 0.045 \pm 0.01$. Evidently the additional
quantum fluctuations due to the presence of loops in the kagome geometry act
to create the same gapless spin-liquid ground state as the nearest-neighbor
model ($J_2 = 0$, Figs.~\ref{F1} and \ref{F2}). Our results are in qualitative
accord with those proposed in Ref.~\cite{ripb} on the basis of VMC studies of
a finite system, although quantitatively the range of stability we deduce is
much narrower.

PESS can be used to calculate further ground-state expectation values.
However, the finite $M(D)$ means that the field-induced magnetization
contains no information useful at zero field. Similarly, finite-$D$
correlation functions have a constant part, which masks the nontrivial
power-law behavior expected of a gapless spin liquid. We have nevertheless
obtained definitive numerical results, in the thermodynamic limit, for the
two key characteristic quantities, $E_0(D,J_2)$ and $M(D,J_2)$. Although our
method is based on gapped states, it is able to indicate its own ``breakdown''
in the event of continuing algebraic convergence \cite{rpvvt}, and thus the
conclusion of a gapless spin-liquid ground state is robust.

To interpret the physical implications of this result, the leading
candidate gapless wave function is the U(1) Dirac-fermion state proposed
in Ref.~\cite{rrhlw}. Although there exist gapless Z$_2$ spin-liquid states
of the KHAF \cite{rlrl}, there is currently neither numerical evidence
\cite{ripb} nor a physical argument in support of these. Heuristically,
gapped spin liquids are favored by the formation of low-energy local
states, such as dimer or plaquette singlets, whereas systems with a net
odd-half-integer spin per simplex do not offer this option. Our results
imply that there is no local unit (such as the hexagon) on the kagome lattice,
and instead the optimal energy is gained by maximizing the kinetic energy of
mobile spinons, leading to the U(1) Dirac-fermion state \cite{rrhlw}, or by
maximizing the contributions from gauge fluctuations \cite{rhfb}. The gapless
spin liquid is expected to have long-ranged entanglement and correlation
functions \cite{rhrlw}, and the U(1) state has no well-characterized
topological order.

Turning to the general question of numerical KHAF studies, our results
constitute a major breakthrough because of the infinite system size.
The fact that all ED and DMRG studies consistently favor gapped states
suggests that systems finite even in only one dimension are not able to
account appropriately for spinon kinetic-energy contributions. Regarding
the question of enforced or emerging spin symmetries, PESS studies enforcing
U(1) \cite{wlpc} or SU(2) \cite{rmchw} symmetry find gapped states with
energies higher than ours (Fig.~\ref{F1}). In our calculations, it is
straightforward to start with a gapped trial PESS wave function and show
that an ordered state of lower energy emerges on projection. In fact all
starting wave functions (symmetric, ordered, arbitrary) lead to the same
final state for a given simplex and update type, with $E_0(D)$ and $M(D)$ as
shown in Figs.~\ref{F1} and \ref{F2}. Thus it appears that symmetry-enforcing
studies are finding excited states, and it is likely that the same applies on
finite systems. Indeed it is argued in Ref.~\cite{ripbc} that a gapped Z$_2$
ground state can lie at lower energy than the gapless U(1) state on a finite
system, but not in the thermodynamic limit. A very recent study using VMC
evaluation of TNS wave functions on finite systems also supports a U(1)
rather than any competing Z$_2$ state \cite{rjkhr}.

Clearly the KHAF is a problem where competing states of very different
character lie very close in energy. We deduce that the large-$N$ approach
offers the best available account of quantum fluctuation effects,
specifically by capturing the kinetic-energy gain of mobile spinons. Our
results also demonstrate the qualitative value of the VMC calculations
\cite{ripb}, which arrive at the gapless spin-liquid ground state by a
different route from PESS, without allowing states of finite $M$. It is
also essential to benchmark whether the PESS Ansatz is ``neutral'' in
its energy accounting, and does not over-emphasize gapless or ordered
states, a question we addressed by comparing the 3- and 9-PESS results
in Figs.~\ref{F1}(a) and \ref{F2}(a).

A further question is whether the algebraic convergence we observe could
cross over to exponential beyond the range of our PESS calculations. If
such a crossover were to begin at $D = 26$, it is hard to argue [consider
Fig.~\ref{F1}(b)] that the difference in extrapolated ground-state energies
could exceed $\Delta E = 0.0001J$. One is then faced with the emergence of an
extremely small energy scale for no apparent reason. This minuscule energy
would have to be the spin gap of the corresponding Z$_2$ state, but clearly
lies far below the DMRG gap. $\Delta E$ lies well below the ``stabilization
energy'' of any of the competing states, whether they arise due to local
resonances, spinon kinetic energy, gauge fluctuations, or any other mechanism.

Turning briefly to experiment, some studies of the material herbertsmithite,
which offers Cu$^{2+}$ ions in an ideal kagome geometry, have indeed suggested
a continuum of fractional spin excitations (Sec.~SI of the SM \cite{sm}).
However, the most recent measurements face competing gapped \cite{rfihl,
rhnwrhbl} and gapless \cite{rhhcnrbl,rkmu} interpretations. In addition, it
remains unclear whether, due to interplane disorder and Dzyaloshinskii-Moriya
interactions, this material is providing a true reflection of kagome physics.

In summary, we have used the method of projected entangled simplex states
to demonstrate that the ground state of the Heisenberg antiferromagnet for
$S = 1/2$ spins on the kagome lattice with only nearest-neighbor interactions
is a gapless quantum spin liquid. A finite next-neighbor interaction reveals
the presence of a narrow regime of gapless spin liquid between states of
finite 120$^{\rm o}$ staggered magnetic order. This spin liquid is thought to
be the U(1) Dirac-fermion state, in which the primary driving force for
spin-liquid behavior is the maximization of spinon kinetic energy.

We thank L. Balents, F. Becca, M. Hermele, and W. Li for helpful discussions.
This work was supported by the National Natural Science Foundation of China
(Grant Nos.~10934008, 10874215, and 11174365), by the National Basic Research
Program of China (Grant Nos.~2012CB921704 and 2011CB309703), and by the
Ministry of Science and Technology of China (Grant No.~2016YFA0300503).

%=============================================================================$
\newpage
\onecolumngrid
\newpage

\section*{Supplemental Material}
%Supplemental Material for: \\
{
\center \bf
Gapless spin-liquid ground state in the $S = 1/2$ kagome
antiferromagnet\\
}
\vspace*{0.2cm}
\begin{center}
H. J. Liao, Z. Y. Xie, J. Chen, Z. Y. Liu, H. D. Xie, R. Z. Huang,
B. Normand, and T. Xiang\\
\vspace*{1.cm}
\end{center}

\twocolumngrid

\setcounter{page}{1}
\setcounter{equation}{0}
\setcounter{figure}{0}
\renewcommand{\theequation}{S\arabic{equation}}
\renewcommand{\thefigure}{S\arabic{figure}}
%=============================================================================$

\subsection*{SI. Kagome Materials and Experiment}

From the standpoint of materials synthesis, great strides have been
made towards a pure $S = 1/2$ kagome system. ZnCu$_3$(OH)$_6$Cl$_2$
(herbertsmithite) \cite{hs} offers a structurally perfect realization
composed of Cu$^{2+}$ ions. Despite the large antiferromagnetic exchange,
$J \simeq 170$-190 K \cite{rrs,rms}, in experiment there is no evidence of
long-ranged magnetic order or even of spin freezing at any temperature down
to 50 mK \cite{rmea,rhea}. There is also no sign of a spin gap in the
excitation spectrum of powder samples \cite{rhea,roea}. Even with the
availability of single crystals \cite{rhhcpsmmnl}, the situation remains
unresolved: neutron spectroscopy \cite{rhhcnrbl} shows a continuum of
fractional excitations and the system was suggested to be gapless (although
the upper limit was set only at $\Delta/J \le 0.1$), but was more recently
interpreted as a gapped state with correlated impurities \cite{rhnwrhbl}.
Similarly, nuclear magnetic resonance studies have been interpreted as showing
a small gap ($\Delta/J \approx 0.05$) \cite{rfihl}, but also as providing no
solid evidence for the presence of one \cite{rkmu}. However, the physics of
herbertsmithite may in fact be controlled by Dzyaloshinskii-Moriya (DM)
interactions \cite{rzea,rcfll,rrmlnm} and, although in-plane defects
\cite{rrmlnm,rdmnm,rldnm,rlkqlhhk,rvkksh,rbnllbdtm} have been excluded
\cite{rfea}, there remain possible in-plane polarization and interplane
coupling effects due to out-of-plane defects. Elsewhere, candidate $S =
1/2$ kagome materials include volborthite \cite{rhhkntkt} and edwardsite
\cite{rioh}, which have distorted lattices, Cu-(1,3bdc) \cite{rnhmn}, which
is ferromagnetic, and vesignieite \cite{royh} and Cd-kapellasite \cite{ro},
which are known to have strong DM interactions, but all raise the pressure for
a theoretical solution to the kagome problem before a definitive experimental
one emerges.

\subsection*{SII. TNS-PESS Calculations}

\subsubsection*{TNS and PESS}

The ability to express a physical system as a tensor network has led to
significant progress in fields as diverse as quantum gravity and quantum
information. The power of tensor manipulation techniques (contraction,
decomposition, and generalized diagonalization) offers new calculational
capabilities for states and operators, as well as new insight into their
physical content. In condensed matter systems, the quantity expressed as a
tensor network is the partition function of a classical system or the wave
function of a quantum system. The TNS construction \cite{Nig,Nishino,Cirac}
is amenable to a renormalization-group approach which allows direct access
to the thermodynamic limit of infinite lattice size, giving TNS methods a
fundamental qualitative advantage over all other numerical approaches to
date.

The TNS wave function respects by construction the area law of entanglement
entropy \cite{AreaLaw}, i.e. its entanglement content is local, which provides
an Ansatz that can be computed in polynomial instead of exponential time. The
truncation parameter is the tensor bond dimension, $D$, which is the number of
auxiliary virtual degrees of freedom \cite{Cirac,XCY+14} introduced on each
lattice bond to describe the local physical degrees of freedom. Heuristically,
$D$ is related to the range of entanglement contained in the wave function
\cite{rpmtm}, with larger $D$ ensuring a more accurate representation. The
conventional formulation of the entangled TNS wave function, PEPS, places
virtual maximally entangled pairs of auxiliary particles on every bond.
However, this pairwise construction does not contain the multipartite
entanglement that is the hallmark of strongly frustrated systems, whence the
requirement for the PESS Ansatz \cite{XCY+14} to allow the full and efficient
application of TNS to this class of problem. We comment in this context that
TNS methods are free of any sign problem \cite{Sign}.

\subsubsection*{Updates, Simplices, and Unit Cells}

The PESS representation is determined by imaginary-time evolution
\cite{Evolution} with an arbitrary starting wave function. In updating
the PESS wave function after each projection step (implemented as a small
time step), one may attempt to include the full system, or bond environment,
in a self-consistent manner by the full-update (FU) scheme \cite{FU}, or
neglect it within the simple-update (SU) approach \cite{JWX08,XCY+14}. SU,
where the bond update corresponds to a (computationally more tractible)
global optimization problem, is exact in 1D and on the Husimi lattice,
where the bonds have no additional environment \cite{BA,LXC+16}. Figures
1(a), 2(a), and 2(b) of the main text compare the values of $E_0(D)$ and
$M(D)$ obtained by the SU and FU techniques. The SU approach clearly
overestimates the expectation value of a local observable, and can be
regarded as providing an upper bound. However, the key information is
that the computational cost of the FU approach prohibits its use beyond
$D = 13$, whereas SU can be employed to at least double this value. Because
the behavior of $E_0(D)$ and $M(D)$ obtained by SU is very similar to the
more efficient FU, and the much greater $D$ delivers far superior estimates
of both [Figs.~1(a), 2(a), and 2(b)], we rely on SU for our primary
conclusions.

Another fundamental, and valuable, degree of freedom within the PESS
construction is the choice of the simplex to use, with larger simplices
delivering more accurate results \cite{XCY+14}. Here we gauge the effects
of the simplex by comparing our 3-PESS results with those from a 9-PESS
Ansatz [Figs.~1(a), 2(a), and 2(b)]. However, when each procedure is optimized,
the larger simplices incur the penalty that the maximum accessible $D$ is
lower. A further variable in our calculations is the size of the unit cell
for which translational invariance is assumed; ordered states and those with
large local bound states may clearly be favored or penalized by different
choices of unit cell. The unit cell can be any multiple of the simplex size,
and we have run our 3-PESS calculations for 3-, 9-, and 12-site unit cells.
Our 9-PESS calculations are made only with a 1-simpex (9-site) unit cell, but
because this accommodates both the $q = 0$ and $\sqrt{3}\times \! \sqrt{3}$
ordering patterns, we believe that no bias is introduced. When the starting
Hamiltonian is extended to the $J_1$--$J_2$ model, this can in principle be
computed in the 3-PESS formalism, but here the $J_2$ bond is an inter-simplex
connection. This makes for a very much less efficient projection and update
procedure, significantly reducing the range of available $D$ values. Thus all
of the $J_1$--$J_2$ results we report were obtained in the 9-PESS framework.

\subsubsection*{Calculation of Expectation Values}

The calculation of a physical expectation value from a PESS representation
with bond dimension $D$ requires in general the full summation over a
two-dimensional reduced tensor network with bond dimension $D^2$. This
becomes the primary bottleneck at larger $D$ in most tensor-network
algorithms \cite{Treview}. We have developed a new approach to this problem,
which is based on the conventional time-evolving block decimation (TEBD)
technique for performing the summation \cite{TEBD}, but avoids calculating
the reduced tensor network by obtaining the expectation values directly
from the local tensors in the PESS representation \cite{DisplaceTEBD}. By
comparison with the TEBD method in two dimensions, the new method reduces
both the computational and memory costs by two orders of $D$, i.e.~from
$D^{10}$ to $D^8$ and from $D^8$ to $D^6$ respectively. Thus we are able to
extend the calculation of some expectation values as far as $D = 30$ ($D =
25$ with full statistical reliability). This extended range is very important
for the kagome Heisenberg antiferromagnet because of the extremely slow
convergence of $E_0(D)$ and $M(D)$ [Figs.~1(a), 1(b), 2(a), and 2(b)].

We gauge the reliability of our results by considering the truncation error
in $E_0(D)$ and $M(D)$ as functions of $D_{\rm mps}$. We comment that the TNS
wave function obtained by repeated TEBD projection is fully variational,
and as such increasing $D$ leads to a lower energy estimate [Fig.~1(a)].
However, the evaluation of physical expectation values is not a variational
procedure, because the MPS basis is changed for each value of $D_{\rm mps}$
and is not determined by minimizing the energy. Thus an increase of
$D_{\rm mps}$ in the evaluation step [Fig.~1(c)] should be considered only
as improving the accuracy of the expectation value being estimated.

\begin{figure}[t]
\begin{center}
\epsfig{file=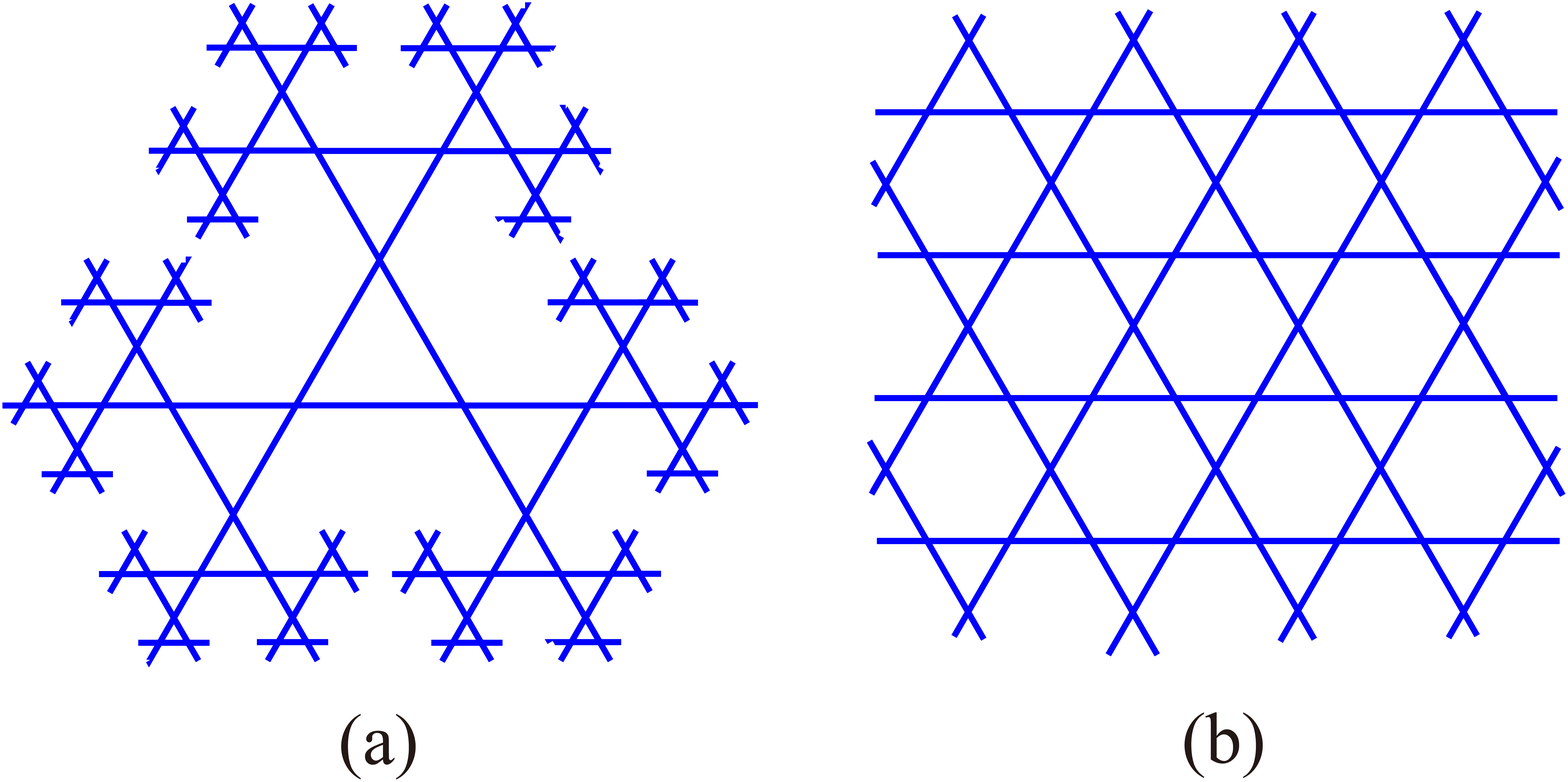,width=8.0cm,angle=0.0}
\end{center}
\caption{(a) Triangular Husimi lattice. (b) Kagome lattice.}
\label{Lattice}
\end{figure}

\subsection*{SIII. On the relationship between Husimi and kagome magnetizations}

While there exists to our knowledge no analytical proof that the magnetization
of the Husimi lattice at finite bond dimension, $M_{\rm H}(D)$, should be an
upper bound on the magnetization of the kagome lattice, $M_{\rm k}(D)$, for the
same value of $D$, here we present a heuristic argument. The triangular Husimi
lattice [Fig.~S1(a)] is a kagome lattice [Fig.~S1(b)] with no longer loops,
i.e. all of the hexagons and longer paths, meaning those with $N > 6$ bonds,
of the kagome lattice are simply absent due to the lack of connectivity of the
Husimi lattice. These loops are mutually frustrating for joint spin-rotation
processes. In a local picture, the short, even-length loops offer extensive
possibilities for gains in resonance energy due to spin fluctuations, and in
a gapless state long loops may also be involved. By contrast, no further energy
is gained if the spins retain their 120-degree configuration, because the two
lattices have equal numbers of bonds per site. Thus the only possible effect
of loop processes on the finite-$D$ magnetization should be to suppress it,
with the result that $M_{\rm H}(D) > M_{\rm k}(D)$. Figures 2(a) and 2(b) of the
main text present a clear numerical demonstration of this inequality.

\end{document}